# Single Atom Catalysts with Halogen Ligands: Elevating the HER Performance of Pd-anchored MoS$_2$ monolayer

Feng Sun, Xuqiang Zhang, Jiangtao Chen, Yun Zhao, Yan Li*

Key Laboratory of Atomic and Molecular Physics & Functional Materials of Gansu Province, College of Physics and Electronic Engineering, Northwest Normal University, Lanzhou 730070, China

**Abstract:**

Single-atom catalysts (SACs) have attracted ever-growing interest due to their high atom-utilization efficiency and potential for cost-effective of hydrogen production. However, enhancing the hydrogen evolution reaction (HER) performance remains a key challenge in developing SACs for HER technology. Herein, we employed first-principles calculations in conjunction with the climbing-image nudged elastic band (CI-NEB) method to explore the effect of surface ligands (F, Cl, Br, I) on the HER performance and mechanism of single-atom (Pd or Cu)-anchored MoS$_2$ monolayer. The results indicate that the relative Gibbs free energy for the adsorbed hydrogen atom in the I-Pd@MoS$_2$ system is an exceptionally low value of -0.13 eV, which is not only comparable to that of Pt-based catalysts but also significantly more favorable than the calculated 0.84 eV for Pd@MoS$_2$. However, the introduction of ligands to Cu@MoS$_2$ deteriorates HER performance due to strong coupling between the absorbed H and ligands. It reveals that the ligand I restructures the local chemical microenvironment surrounding the SAC Pd, leading to impurity bands near the Fermi level that couple favorably with the s states of H atoms, yielding numerous highly active sites to enhance catalytic performance. Furthermore, the CI-NEB method elucidates that the enhanced HER mechanism for the I-Pd@MoS$_2$ catalyst should belong to the coexistence of the Volmer-Tafel and Volmer-Heyrovsky reactions. This investigation provides a valuable framework for the experimental design and development of innovative single-atom catalysts.

* Corresponding author: Yan Li (liyan-nwnu@163.com)

# I. INTRODUCTION

Facing the current global energy crisis, the pursuit of carbon-neutral, clean, and renewable energy alternatives is imperative. Hydrogen ($H_2$), with high energy density and non-polluting nature, has attracted extensive attention and is considered as a leading candidate for energy carrier and storage [1-3]. Multiple hydrogen storage and production technologies have been proposed [4, 5], such as electrochemical water splitting and fossil fuel calcination. Among these, electrochemical water splitting is particularly prominent as it can transform electrical energy into hydrogen fuel with relatively low environmental impact, high safety, and cost-effectiveness [6-12]. The central challenge in advancing electrochemical hydrogen evolution technology lies in enhancing the efficiency of the hydrogen evolution reaction (HER), which is significantly influenced by the choice of catalysts. Substantial efforts have been dedicated to the construction and improvement of catalysts, covering a range from bulk to low dimension and from single-cluster catalysts (SCCs) to single-atom catalysts (SACs). SACs have emerged as a new research frontier in HER research [13], combining the high activity of homogeneous catalysts with the good stability of heterogeneous ones [14]. Metal SACs, in particular, can achieve 100% metal utilization and offer broad tunability, making them an attractive candidates for enhancing HER performance [15]. For instance, anchoring Pt atoms onto graphene nanosheets could effectively improve HER performance by up to 37 times compared to commercial Pt/C [16]. Also, single-atom Pt doping can activate the otherwise inert 2D $MoS_2$ surface for HER [17]. The strong interaction between single metal atoms and their supports can reconstruct the atomic and electronic structures, facilitating electron transfer in HER reactions and ultimately enhancing HER performance. Therefore, the microenvironment around the catalysts has a crucial impact on HER performance, leading to numerous researches on regulating the coordination environment around SACs to improve HER performance.

The regulating methods for SACs can generally be divided into two common types: in-plane and out-of-plane modification. For example, N doing within the plane of graphene substrate has been shown to significantly enhance the HER performance and stability of single Co atoms, with a very low overpotential of 30 mV observed [18]. The excellent HER activity is related to single Co centers coordinated to doped N, and the great stability is due to the strong interaction between Co and N atoms caused by high-temperature treatment. Similarly, introducing axial ligands onto the out-of-plane of NiFe-layered double hydroxide nanoarrays can largely promote the alkaline HER performance of the Pt SACs because the largest first electron affinity of the axial ligand changes the electronic states of the Pt SACs, as verified by first principles calculations showing the change in the average energy levels of the d orbitals of Pt atoms [19]. These results highlight the importance of the catalyst's microenvironment in determining its HER performance.

Despite the above advances, there is still limited understanding of the impact of ligands on the HER performance of SACs. In this work, the first-principles calculations combined with the climbing-image nudges elastic band (CI-NEB) method is utilized to investigate the effects of halogen ligands (F, Cl, Br and I) on the atomic and electronic structures, as well as the HER performance of the SACs (Pd or Cu) supported on the $MoS_2$ monolayer. This study aims to provide insights into the role of ligands in modulating the catalytic activity of SACs, thereby contributing to the development of more efficient HER catalysts.

**II. COMPUTATIONAL METHODS**

For this study, we utilized the Vienna ab initio simulation package (VASP) [20, 21] with the projector augmented wave (PAW) approach [22] for geometry optimizations and calculations of total energies and electronic structures. The exchange-correlation functionals were described by the generalized gradient approximation of Perdew-Burke-Ernzerhof (GGA-PBE) [23]. To construct a 2D surface structure, a vacuum layer of 20 Å along the *c*-axis was employed to eliminate periodic interactions. The fixed *c*-axis method is adopted to acquire the lowest energy configurations, and all atomic positions are thoroughly relaxed with an energy convergence threshold set at $10^{-5}$ eV in the optimization and electronic structure calculations. Since the force convergence threshold depends on the initial crystal structure, a value of 0.005 eV/Å is applicable for the metal-supported $MoS_2$ monolayer and metal-ligand-supported monolayer. In the case of the metal-ligand-supported monolayer with adsorbed hydrogen, a force convergence threshold lower than 0.05 eV/Å is adopted. The kinetic energy cutoff for the plane-wave basis is set to 400 eV, and a *k*-mesh of 5×5×1 within the Brillouin zone is utilized. To screen the minimum energy pathways and explore the expansion behavior of surface atoms, the climbing image nudged elastic band (CI-NEB) method was applied [24]. In this implementation, 6 intermediate images were incorporated, along with a force convergence threshold of 0.08 eV/Å and a *k*-mesh of 3×3×1.

## III. RESULTS AND DISCUSSIONS

### 1. Crystal structures

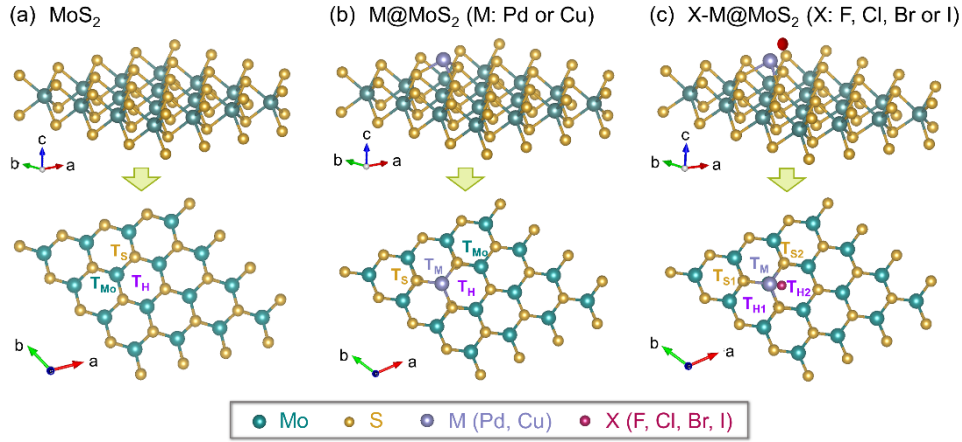

**Fig. 1.** Schematic illustrations of the monolayers of (a) $MoS_2$, (b) M@$MoS_2$ (M: Pd or Cu) and (c) X-M@$MoS_2$ (X: F, Cl, Br or I) with the initial position markings of M and X atoms. $T_S$, $T_{Mo}$, $T_M$ and $T_H$ denote the top positions of S, Mo, Pd atoms and the hollow center, respectively. $T_{H1}$ and $T_{H2}$, represent two nonequivalent top positions of the hollow centers. $T_{S1}$ and $T_{S2}$ signify two nonequivalent top positions of the S atoms.

To initiate the investigation into the structural characteristics and stability of the systems, three distinct initial configurations are systematically constructed. In these configurations, the single atom catalysts (SACs) are positioned atop the hollow site, the tops the S atom, and the Mo atom, which are hereafter denoted as $T_H$, $T_S$, and $T_{Mo}$, respectively, as illustrated in Fig. 1(a). The primary objective of this construction is to identify the most stable structural arrangement among them. Our comprehensive computational analyses yielded conclusive results indicating that the SACs exhibited a distinct preference for being supported upon the Mo atom site. Specifically, the $T_{Mo}$ configuration was found to possess the lowest energy state (see Table S1 in Supporting Information), thereby rendering it the most energetically favorable configuration. This finding is of fundamental significance as it provides crucial insights into the preferred interaction geometry between the SACs and the $MoS_2$ support. The computational analysis indicates that the SACs exhibit a preference for being supported on the top of Mo, with the $T_{Mo}$ configuration the most energetically favorable. As clearly shown in Fig. 2a, for the optimized Pd@$MoS_2$, the bond lengths of $D_{Pd-S}$ and $D_{Pd-Mo}$ are 2.33 Å and 2.81 Å, respectively, while in Cu@$MoS_2$, the $D_{Cu-S}$ and $D_{Cu-Mo}$ are 2.25 Å and 2.84 Å. Although the different supported SAC has a slight influence on these bond lengths, it appears to have a more pronounced effect on the S-S bond lengths surrounding the Pd or Cu atoms. In Cu@$MoS_2$, the S-S bond length is 3.26 Å, increasing to 3.39 Å in Pd@$MoS_2$, likely due to the larger atomic radius of Pd compared to Cu. Nevertheless, the S-S bond lengths away from the SACs remain comparable to those in pristine $MoS_2$ monolayer (about 3.18 Å). This implies that SACs primarily affect the structural reconstruction of nearby atoms, as evidenced by the minimal differences in lattice parameters between M@$MoS_2$ and pure $MoS_2$. Additionally, the S-Mo bonds are relatively unaffected by the presence of SACs.

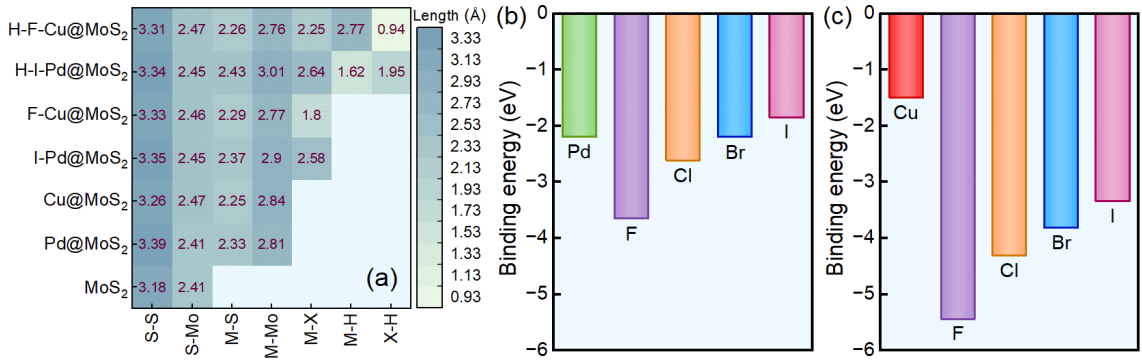

Fig. 2 (a) Bond lengths (Å) of S-S, S-Mo, M-S, M-Mo, M-X, M-H and X-H in seven different systems. X denotes I element in I-Pd@$MoS_2$ and H-I-Pd@$MoS_2$, F element in F-Cu@$MoS_2$ and H-F-Cu@$MoS_2$. M represents Pd in Pd@$MoS_2$, I-Pd@$MoS_2$ and H-I-Pd@$MoS_2$, while indicates Cu in Cu@$MoS_2$, F-Cu@$MoS_2$ and H-F-Cu@$MoS_2$. Binding energies of transition metal M (M = Pd or Cu) and halogen element X (X = F, Cl, Br and I) in the structures containing (b) Pd or and (c) Cu.

Four different ligands (F, Cl, Br and I) are introduced around SACs in the above

M@MoS$_2$ (M = Pd or Cu), with the corresponding structures referred to X-M@MoS$_2$ (X = F, Cl, Br or I). Each structure, with four distinct initial configurations with the halogen atom positioned at the T$_H$, T$_S$, T$_{Mo}$ and T$_M$ sites (as shwon in Fig. 1b), was fully relaxed. Our findings indicate that the introduction of halogen elements has a relatively minor influence on the local structure. The S-S bond length decreases from 3.39 Å in Pd@MoS$_2$ to 3.35 Å in I-Pd@MoS$_2$, and increases from 3.26 Å in Cu@MoS$_2$ to 3.33 Å in F-Cu@MoS$_2$. However, the lengths of the S-Mo and M-Mo bonds increase in I-Pd@MoS$_2$ and decrease in F-Cu@MoS$_2$, while the M-S bond length increases in both I-Pd@MoS$_2$ and F-Cu@MoS$_2$. As known, the atom radius of halogens (R$_X$) follows the trend R$_F$ < R$_{Cl}$ < R$_{Br}$ < R$_I$, thus the bond lengths of D$_{X-Pd}$ (X = F, Cl, Br, I) between the halogen stoms (X = F, Cl, Br, I) and the Pd atoms are 1.96, 2.38, 2.4 and 2.58 Å, respectively. Comparatively, the Cu radius is smaller than that of Pd, the bond lengths of D$_{X-Cu}$ exhibit shorter values of 1.80, 2.11, 2.25 and 2.43 Å.

The binding energies ($E_{bX}$ and $E_{bM}$) of X atom in X-M@MoS$_2$ and M atom in M@MoS$_2$ were calculated using the following expression [25, 26]

$$E_{bX} = E_{X-M-pure} - E_{M-pure} - E_X$$
$$E_{bM} = E_{M-pure} - E_{pure} - E_M$$

where $E_{X-M-pure}$, $E_{M-pure}$ and $E_{pure}$ are the total energies of X-M@MoS$_2$, M@MoS$_2$ and pure MoS$_2$ monolayer, and $E_X$ and $E_M$ are the total energies of isolated X and M atoms, respectively. In fact, their binding energies are directly proportional to the X-M bond lengths. Moreover, the binding energies of Pd-MoS$_2$ and Cu-MoS$_2$ are negative, indicating that the bonding of Pd and Cu atoms to MoS$_2$ monolayer belongs to the exothermic reactions, thereby supporting the stabilities. The $E_{bM}$ value of Cu is slightly greater than that of Pd bonding to MoS$_2$, however, the $E_{bX}$ values of halogen atoms bonding to Cu@MoS$_2$ are less than those bonding to Pd@MoS$_2$. This is attributed to different electron configurations of [Ar]4s$^1$3d$^{10}$ for Cu and [Kr] 4d$^{10}$ for Pd, which leads the coupling between s and d electrons in X-Pd bonds, the coupling between s and s electrons in X-Cu bonds due to the closed shell d$^{10}$.

In the stable structure given in Fig. 3a-b, the I-Pd bond in I-Pd@MoS$_2$ is not perpendicular to the surface of MoS$_2$, however, the F-Cu bond in F-Cu@MoS$_2$ is completely perpendicular to the MoS$_2$ surface. Therefore, the ligands and SACs exhibit significant interactions that substantially influence the local atomic structure and energetic properties of the monolayer MoS$_2$ surface. Therefore, they would play a crucial role in enhancing the HER activity, despite having minimal impact on the lattice parameters.

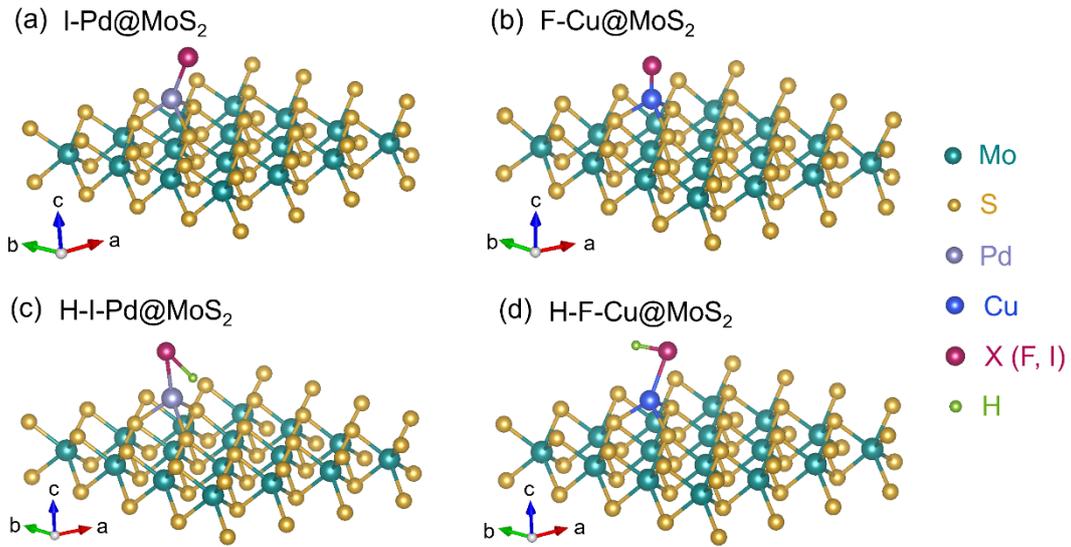

**Fig. 3.** Sable structures of (a) I-Pd-MoS$_2$, (b) F-Cu@MoS$_2$, (c) H-I-Pd@MoS$_2$ and (d) H-F-Cu@MoS$_2$.

Furthermore, five initial sites of H atom, containing two T$_S$ sites (T$_{S1}$ and T$_{S2}$), two T$_H$ sites (T$_{H1}$ and T$_{H2}$) and one T$_M$ site (Fig. 1c) to analyze the HER activity. The calculated results demonstrate a significant sensitivity of the final structures to the initial configurations. As presented in Fig. 3c-d, the stable H-I-Pd@MoS$_2$ structure from the initial T$_{H2}$ configuration shows both I and Pd atoms on top of the Mo atom, with the I-Pd bond almost perpendicular to the MoS$_2$ surfaces. The lengths of H-Pd and H-I bonds are 1.62 and 1.95 Å, respectively, which are shorter than H-Cu and H-F lengths (2.77 and 0.94 Å). Clearly, the HER performance of I-Pd@MoS$_2$ and F-Cu@MoS$_2$ primarily depends on the SAC Pd and ligand F, respectively. Typically, the H-F bond exhibits strong polarity and covalent characteristic which hinders the desorption of H atom, being supported by the large relative Gibbs free energy noted in F-Cu@MoS$_2$, as discussed below.

## 2. Electronic structure

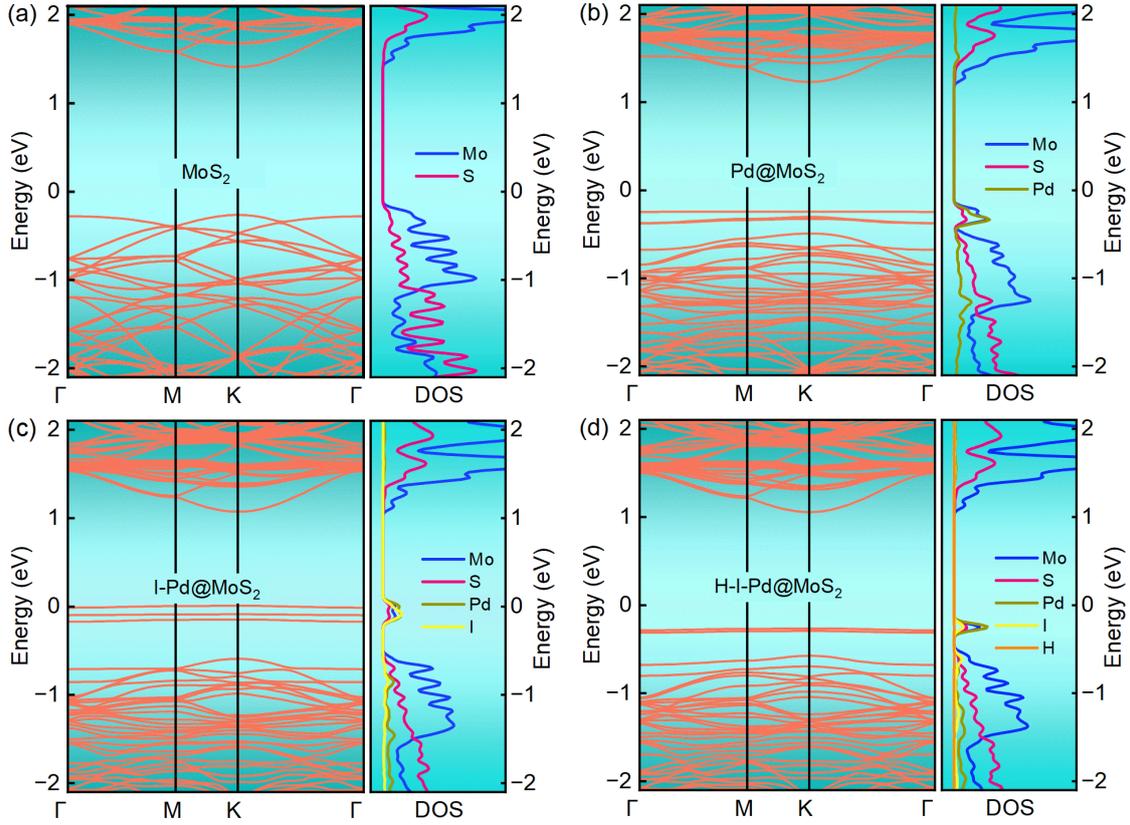

**Fig. 4.** Band structures and DOS plots of (a) $MoS_2$, (b) $Pd@MoS_2$, (c) $I-Pd@MoS_2$ and (d) $H-I-Pd@MoS_2$ systems.

Fig. 4 analyzes the calculated band structure and DOS results. In Fig. 4a, both the valence-band maximum (VBM) and conduction-band minimum (CBM) of $MoS_2$ monolayer are located at K point, leading to a direct bandgap of 1.69 eV. The DOS demonstrates the contribution of Mo atoms to CBM and VBM is obviously greater than that of S atoms. As shown in Fig. 4b, the introduced Pd in Pd@MoS2 SAC creates three impurity levels below the Fermi energy level, which is mainly attributed to the hybridization caused by the strong interaction between d electrons of the Pd and Mo atoms. The addition of ligand I element, as seen in Fig. 4c, could weaken the d-d hybridization between Pd and Mo atoms, but the p-d hybridization between I and Pd atoms dominates the three impurity levels. This causes the impurity levels to be lifted across the Fermi surface. Comparatively, the S element has a minor influence on the impurity levels

As known, the Pt element has highly HER activities due to the d electrons near the Fermi level. Normally, the d-band center theory is used to evaluate the relations between the electrocatalytic activity of transition metals and the energy level of d electron [27, 28], which is defined as [29, 30]

$$\varepsilon_d = \frac{\int_{-\infty}^{+\infty} ED(E)dE}{\int_{-\infty}^{+\infty} D(E)dE}$$

where $E$ and $D(E)$ are the energy of d electron and its corresponding DOS. The $\varepsilon_d$ values of Mo and Pd atoms in I-Pd@MoS$_2$ are -0.80 and -2.05 eV, and change to -0.83 and -2.01 eV upon H absorption. Clearly, these changes result from the variations in the d-states of the impurity bands. In F-Cu@MoS$_2$ structure, the $\varepsilon_d$ values of Mo and Cu atoms are -1.18 and -1.87 eV, respectively, but largely decrease to -2.05 and -3.58 eV in H-F-Cu-Pd@MoS$_2$. These results suggest that despite the significant changes in the energy distribution, the d electrons in I-Pd@MoS$_2$ remain relatively stable. Indeed, subsequent findings indicate that this stable d-state is advantageous for enhancing hydrogen evolution performance.

## 3. HER property

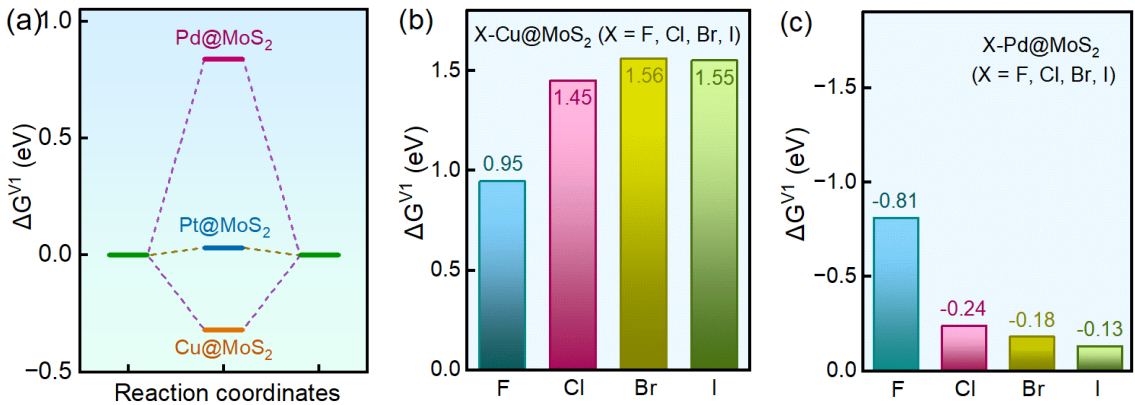

**Fig. 5.** (a) Reaction free energy landscape for HER in MoS$_2$ monolayers with supported SACs Pd, Pt and Cu. Comparative depiction of the relative Gibbs free energies for (b) X-Cu@MoS$_2$ (c) and X-Pd@MoS$_2$ (X = F, Cl, Br, I).

HER can be adequately explained by either the Volmer-Heyrovsky or Volmer-Tafel mechanisms, or a combination of both [31, 32]. The Volmer reaction, which involves the adsorption of hydrogen, is widely recognized as the initial and critical step in HER, fundamentally determining the feasibility of subsequent reactions. Herein, we utilize the relative free energy of the Volmer reaction ($\Delta E^{Vn}$) to preliminarily evaluate the feasibility of the reaction [33, 34].

$$\Delta E^{Vn} = E^n - E^{n-1} - \frac{1}{2}E^{H_2} + \Delta E_{vib} - T\Delta S$$

where $E_n$ and $E_{n-1}$ are the total energies of X-M@MoS$_2$ (X = F, Cl, Br or I; M = Pd or Cu) with $n$ and $n$-1 H atoms absorbed on the surface, $E^{H_2}$ is the total energy of one hydrogen molecule, $\Delta E_{vib}$ is the difference in vibration energies, and $\Delta S$ is the difference in entropy, which could also be defined as

$$\Delta E_{vib} = E_{vib}^n - E_{vib}^{n-1} - \frac{1}{2}E_{vib}^{H_2}$$

$$\Delta S = S_{vib}^n - S_{vib}^{n-1} - \frac{1}{2}S_{vib}^{H_2}$$

Fig. 5a illustrates that the $\Delta G^{V1}$ values for SACs of Pt, Pd and Cu in the absence of ligands are 0.07, 0.84 and -0.32 eV, respectively. It is widely accepted that a $\Delta G^{V1}$ value approaching zero is beneficial for HER, as it facilitates both hydrogen adsorption and desorption. Therefore, Pt SACs should exhibit the most excellent HER performance, while Pd and Cu SACs present significant potential for enhancing HER performance through further optimization. The positive and negative values indicate the absorption and release of energy during the hydrogen adsorption process. Thus, SACs Pd exhibits a tendency towards facile desorption of H atoms, whereas Cu demonstrates an affinity for H adsorption.

In Fig. 5b, the introduction of the halogen ligands F, Cl, Br and I into Cu@MoS$_2$ shifts the $\Delta G^{V1}$ values from negative to positive, increasing them to 0.95, 1.45, 1.56 and 1.55 eV, respectively. These values are much higher than those of pristine Cu@MoS$_2$, and further deviate from zero. Resultantly, M-Cu@MoS$_2$ is not a suitable candidate for HER materials. In M-Pd@MoS$_2$, the halogen ligands (M=F, Cl, Br and I) reduce the $\Delta G^{V1}$ values from +0.84 eV to -0.81, -0.24, -0.18 and -0.13 eV, respectively. Obviously, the halogen ligands Cl, Br and I can effectively enhance the HER performance by lowering the $\Delta G^{V1}$ values into the negative range. However, the highest electronegativity of F among the four ligands is not conducive to enhancing the HER performance of Pd@MoS$_2$.

This implies that excessive coupling between the ligand and the d electrons of Pd is detrimental to forming appropriate electronic structures for efficient HER. Furthermore, our findings indicate that the $\Delta G^{V1}$ values of the adjacent adsorption sites surrounding the Pd and ligand (Cl, Br or I) atoms have significantly decreased, approaching zero. This suggests the presence of multiple active sites for HER.

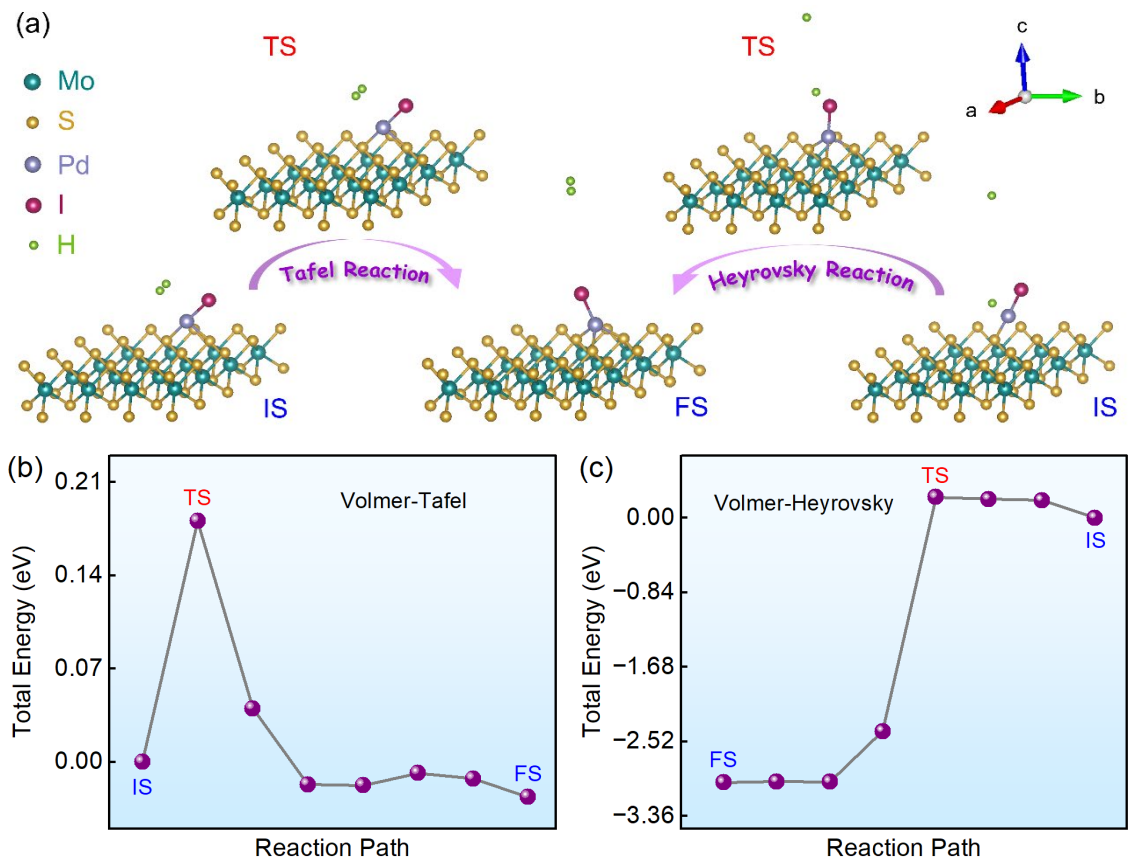

**Fig. 6.** Stable structures of the initial, transition, final states in (a) Volmer-Tafel reaction and Volmer-Heyrovsky reaction, in which the same final state is employed for the two reactions. Energy profiles for (b) Volmer-Tafel reaction and (c) Volmer-Heyrovsky reaction.

Fig. 6 further explores the HER mechanism of I-Pd@MoS$_2$ using the CI-NEB method. As illustrated in Fig. 6a, in the Tafel reaction, two absorbed H atoms on the surface pass through a transition state (TS) and then combine to form a hydrogen molecule. Structurally, the desorption of the H atoms during the Tafel reaction exerts a minimal influence on the bond lengths of Pd-I and Pd-S, altering only the bond direction. In the TS, two H atoms have tended to bond to form hydrogen molecule with an H-H bond length 0.75 Å, which is notably shorter than 0.80 Å observed in the initial state (IS). In the final state (FS), the hydrogen molecule is nearly perpendicular to the surface. The effect of altering the orientation of hydrogen molecules on energy is counteracted by the weakened interaction between the surface and the hydrogen molecule due to the increased distance. Thus, the total energy difference between the IS and FS is minimal (~ 0.03 eV).

The Tafel reaction path reveals a transition state with slightly elevated energy between the IS and FS. The energy difference between this TS and IS is 0.19 eV, which is lower than the $\Delta E^{V2}$ value of 0.56 eV. This indicates that the feasibility of the Volmer-Tafel reaction is contingent upon the successful occurrence of the second Volmer reaction. More intriguingly, in the Heyrovsky reaction, the FS transitions to the TS with a comparable reaction barrier of 0.23 eV, similar to that observed in the Tafel reaction. Therefore, the HER mechanism of I-Pd@$MoS_2$ involves a concurrent occurrence of both the Volmer-Tafel and Volmer-Heyrovsky reactions.

## 4. SUMMARY

In summary, the effects of surface ligands (F, Cl, Br, I) on the HER performance and mechanism of single-atom (Pd, Cu)-anchored $MoS_2$ are systematically investigated by using the first-principles calculation combined with the CI-NEB method. It is found that halogen ligands can effectively enhance the performance of the SAC Pd, especially the ligand I can reduce the relative Gibbs free energy to almost zero, which is comparable to the Pt-based catalyst. However, the ligands appear to have limited impact on enhancing the HER performance of SAC Cu. In contrast, the introduction of ligand I to Pd@$MoS_2$ reconstructs the microenvironment surrounding the Pd atom and modifies its electronic structure, thereby significantly improving the HER performance. In I-Pd@$MoS_2$, three impurity bands primarily composed of the d-states of Pd and Mo atoms are located near the Fermi level. These d-states exhibit relative stability and favorable coupling with the s-states of H atom, resulting in minimal impact on the energy levels. Consequently, this leads to a relative Gibbs free energy that approaches zero. Moreover, the HER mechanism of I-Pd@$MoS_2$ is confirmed to be a coexistence of Volmer-Tafel and Volmer-Heyrovsky reactions.

AUTHOR'S CONTRIBUTION

All authors contributed equally to this work.

ACKNOWLEDGMENT:

DATA AVAILABILITY

The data that supports the findings of this study are available within the article and its supplementary material.

**References:**

[1] El-Adawy M, Dalha IB, Ismael MA, Al-Absi ZA, Nemitallah MA. Review of sustainable hydrogen energy processes: production, storage, transportation, and color-coded classifications. Energ Fuel 2024;38:22686-718.
[2] Yin H, Rong F, Xie Y. A review of typical transition metal phosphides electrocatalysts for hydrogen evolution reaction. Int J Hydrogen Energ 2024;52:350-75.
[3] Guo W, Wang Z, Wang X, Wu Y. General design concept for single-atom catalysts toward heterogeneous catalysis. Adv Mater 2021;33:2004287.
[4] Ma N, Zhao W, Wang W, Li X, Zhou H. Large scale of green hydrogen storage: Opportunities and challenges. Int J Hydrogen Energ 2024;50:379-96.
[5] Song K, Zhang H, Lin Z, Wang Z, Zhang L, Shi X, et al. Interfacial engineering of cobalt thiophosphate with strain effect and modulated electron structure for boosting electrocatalytic hydrogen evolution reaction. Adv Funct Mater 2024;34:2312672.
[6] Tang Y, Yang C, Xu X, Kang Y, Henzie J, Que W, et al. MXene nanoarchitectonics: defect-engineered 2D MXenes towards enhanced electrochemical water splitting. Adv Energy Mater 2022;12:2103867.
[7] Su H, Song S, Li N, Gao Y, Li P, Ge L, et al. Flexibility tuning of dual-metal S-Fe-Co-$N_5$ catalysts with O-axial ligand structure for electrocatalytic water splitting. Adv Energy Mater 2023;13:2301547.
[8] Wang Y, Wang S, Ma Z-L, Yan L-T, Zhao X-B, Xue Y-Y, et al. Competitive coordination-oriented monodispersed ruthenium sites in conductive MOF/LDH hetero-nanotree catalysts for efficient overall water splitting in alkaline media. Adv Mater 2022;34:2107488.
[9] Gao T, Tang X, Li X, Wu S, Yu S, Li P, et al. Understanding the atomic and defective interface effect on ruthenium clusters for the hydrogen evolution reaction. ACS Catal 2023;13:49-59.

[10] Li X, Mitchell S, Fang Y, Li J, Perez-Ramirez J, Lu J. Advances in heterogeneous single-cluster catalysis. Nat Rev Chem 2023;7:754-67.

[11] Tian Y, Luo Y, Wu T, Quan X, Li W, Wei G, et al. Coupling interaction between precisely located Pt single-atoms/clusters and NiCo-layered double oxide to boost hydrogen evolution reaction. Adv Funct Mater 2024;34:2405919.

[12] Lyu F, Zeng S, Jia Z, Ma F-X, Sun L, Cheng L, et al. Two-dimensional mineral hydrogel-derived single atoms-anchored heterostructures for ultrastable hydrogen evolution. Nat Commu 2022;13:6249.

[13] Busari FK, Babar ZUD, Raza A, Li G. Unlocking new frontiers: Boosting up electrochemical catalysis with metal clusters and single-atoms. Sustainable Materials and Technologies 2024;40:e00958.

[14] Niu H, Wang Q, Huang C, Zhang M, Yan Y, Liu T, et al. Noble metal-based heterogeneous catalysts for electrochemical hydrogen evolution reaction. Applied Sciences 2023;13:2177.

[15] Alam N, Noor T, Iqbal N. Catalyzing sustainable water splitting with single atom catalysts: recent advances. Chem Rec 2024;24:e202300330.

[16] Cheng N, Stambula S, Wang D, Banis MN, Liu J, Riese A, et al. Platinum single-atom and cluster catalysis of the hydrogen evolution reaction. Nat Commu 2016;7:13638.

[17] Deng J, Li H, Xiao J, Tu Y, Deng D, Yang H, et al. Triggering the electrocatalytic hydrogen evolution activity of the inert two-dimensional $MoS_2$ surface via single-atom metal doping. Energy Environ Sci 2015;8:1594-601.

[18] Fei H, Dong J, Arellano-Jiménez MJ, Ye G, Dong Kim N, Samuel ELG, et al. Atomic cobalt on nitrogen-doped graphene for hydrogen generation. Nat Commu 2015;6:8668.

[19] Zhang T, Jin J, Chen J, Fang Y, Han X, Chen J, et al. Pinpointing the axial ligand effect on platinum single-atom-catalyst towards efficient alkaline hydrogen evolution reaction. Nat Commu 2022;13:6875.

[20] Kresse G, Furthmüller J. Efficient iterative schemes for ab initio total-energy calculations using a plane-wave basis set. Phys Rev B 1996;54:11169-86.

[21] Kresse G, Joubert D. From ultrasoft pseudopotentials to the projector augmented-wave method. Phys Rev B 1999;59:1758-75.

[22] Blöchl PE. Projector augmented-wave method. Phys Rev B 1994;50:17953-79.

[23] Perdew JP, Burke K, Ernzerhof M. Generalized gradient approximation made simple. Phys Rev Lett 1996;77:3865-8.

[24] Henkelman G, Uberuaga BP, Jónsson H. A climbing image nudged elastic band method for finding saddle points and minimum energy paths. The Journal of Chemical Physics 2000;113:9901-4.

[25] Cheng YC, Zhu ZY, Mi WB, Guo ZB, Schwingenschlögl U. Prediction of two-dimensional diluted magnetic semiconductors: Doped monolayer $MoS_2$ systems. Phys Rev B 2013;87:100401.

[26] Ren Y, Sun X, Qi K, Zhao Z. Single atom supported on $MoS_2$ as efficient electrocatalysts for the $CO_2$ reduction reaction: A DFT study. Appl Surf Sci 2022;602:154211.

[27] Hammer B, Nørskov JK. Electronic factors determining the reactivity of metal surfaces. Surf Sci 1995;343:211-20.

[28] Hammer B, Morikawa Y, Nørskov JK. CO chemisorption at metal surfaces and overlayers. Phys Rev Lett 1996;76:2141-4.

[29] Ouyang Y, Ling C, Chen Q, Wang Z, Shi L, Wang J. Activating inert basal planes of MoS$_2$ for hydrogen evolution reaction through the formation of different intrinsic defects. Chem Mater 2016;28:4390-6.

[30] Shu H, Zhou D, Li F, Cao D, Chen X. Defect engineering in MoSe$_2$ for the hydrogen evolution reaction: from point defects to edges. ACS Appl Mater Interfaces 2017;9:42688-98.

[31] Zhang L, Zheng Y, Wang J, Geng Y, Zhang B, He J, et al. Ni/Mo bimetallic-oxide-derived heterointerface-rich sulfide nanosheets with Co-doping for efficient alkaline hydrogen evolution by boosting volmer Reaction. Small 2021;17:2006730.

[32] Zhu J, Hu L, Zhao P, Lee LYS, Wong K-Y. Recent advances in electrocatalytic hydrogen evolution using nanoparticles. Chem Rev 2020;120:851-918.

[33] Zhang Y, Lei H, Duan D, Villota E, Liu C, Ruan R. New Insight into the mechanism of the hydrogen evolution reaction on MoP(001) from first principles. ACS Appl Mater Interfaces 2018;10:20429-39.

[34] Ling C, Shi L, Ouyang Y, Chen Q, Wang J. Transition metal-promoted V$_2$CO$_2$ (MXenes): A New and highly active catalyst for hydrogen evolution reaction. Adv Sci 2016;3:1600180.